\documentclass{pasj00}
\SetRunningHead{T. Suenaga et al.}{Spectroscopy of brown dwarfs in Orion}
\Received{2013/01/15}
\Accepted{2013/10/30}
\Published{}

\usepackage{txfonts}
\usepackage{natbib}
\usepackage{graphicx}
\usepackage{url}
\newcommand{\mm}{\ensuremath{\mu {\rm m}}}
\newcommand{\plms}{\ensuremath{\pm}}
\newcommand{\teff}{\ensuremath{T_{\rm eff}}}
\newcommand{\av}{\ensuremath{A_V}}

\DeclareAbbreviation\pasa{PASA}

\begin{document}
\title{Multi-Object and Long-Slit Spectroscopy of Very Low Mass Brown Dwarfs in the Orion Nebular Cluster}
\author{
	Takuya \textsc{Suenaga}, \altaffilmark{1}
	Motohide \textsc{Tamura}, \altaffilmark{1,2,3}
	Masayuki \textsc{Kuzuhara}, \altaffilmark{4}\\
	Kenshi \textsc{Yanagisawa}, \altaffilmark{5}
	Miki \textsc{Ishii}, \altaffilmark{6}
	and
	Philip W. \textsc{Lucas}, \altaffilmark{7}
}

\altaffiltext{1}{Department of Astronomical Science, The Graduate University for Advanced Studies (Sokendai),\\ 2-21-1 Osawa, Mitaka, Tokyo, 181-8588}
\altaffiltext{2}{National Astronomical Observatory of Japan, 2-21-1 Osawa, Mitaka, Tokyo 181-8588}
\altaffiltext{3}{Department of Astronomy, The University of Tokyo, 7-3-1, Hongo, Bunkyo-ku, Tokyo, 113-0033}
\altaffiltext{4}{Department of Earth and Planetary Sciences, Tokyo Institute of Technology, 2-21-1 Ookayama, Meguro-ku, Tokyo 152-8551}
\altaffiltext{5}{Okayama Astronomical Observatory, National Astronomical Observatory of Japan,\\ 3037-5 Honjo, Kamogata, Asaguchi, Okayama, 719-0232}
\altaffiltext{6}{Subaru Telescope, National Astronomical Observatory of Japan, 650 North A'ohoku place, Hilo, HI, 96720, USA}
\altaffiltext{7}{Centre for Astrophysics Research, University of Hertfordshire, College Lane, Hatfield AL10 9AB}
\email{takuya.suenaga@nao.ac.jp}


\maketitle

\begin{abstract}
We present the results of a $H-$ and $K-$band multi-object and long-slit spectroscopic survey of substellar mass candidates in the outer regions of the Orion Nebula Cluster. The spectra were obtained using MOIRCS on the 8.2-m Subaru telescope and ISLE on the 1.88-m telescope of Okayama Astronomical Observatory. Eight out of twelve spectra show strong water absorptions and we confirm that their effective temperatures are $\leq $ 3000 K (spectral type $\geq $ M6) from a chi-square fit to synthetic spectra. We plot our sources on an HR diagram overlaid with theoretical isochrones of low-mass objects and identify three new young brown dwarf candidates. One of the three new candidates is a cool object near the brown dwarf and planetary mass boundary. Based on our observations and those of previous studies, we determine the stellar ($0.08<M/M_{\odot}<1$) to substellar ($0.03<M/M_{\odot}<0.08$) mass number ratio in the outer regions of the Orion nebular cluster to be $3.5\pm 0.8$. In combination with the number ratio reported for the central region ($3.3^{+0.8} _{-0.7}$), this result suggests the number ratio does not simply change with the distance from the center of the Orion nebular cluster.
\end{abstract}

\section{Introduction}
Brown dwarfs (BDs) are substellar mass objects between planetary and stellar mass regimes ($0.013\leq M/M_{\odot}\leq 0.075$). Since their discovery in the 1990s, hundreds of examples have been reported, and NASA's WISE mission has recently added another hundred BDs \citep{2011ApJS..197...19K}. However, since the typical thermal Jeans mass is $\sim 1\ M_{\odot}$ in molecular cloud cores \citep[][and references therein]{Larson:1999lr}, the gravitational collapse might be expected to form stars, not BDs. Therefore the formation mechanism of BDs is still unclear. Several theories for the formation of BDs have been suggested,  including turbulent fragmentation \citep{2004ApJ...617..559P,2005A&A...430.1059B}, magnetic field confinement \citep{2001ApJ...551L.167B}, stellar embryo ejection through dynamical interactions \citep{2001AJ....122..432R,2002MNRAS.332L..65B}, and photo-evaporation of embryos by nearby hot stars \citep{2004A&A...427..299W}, though none of these theories have yet proven conclusive. Star forming theories predict the existence of even planetary mass objects (PMOs) with a lowest mass limit of 0.001 -- 0.010 $M_{\odot}$ related to the opacity of the contracting object, known as opacity limited fragmentation \citep[e.g.,][]{1976MNRAS.176..367L,1976MNRAS.176..483R,2001ApJ...551L.167B,2006A&A...458..817W}. However, since they are not massive enough to sustain hydrogen burning and become fainter with time, substellar objects are very faint in the field. Therefore, only a few PMOs have been reported in the field \citep[e.g.,][]{Cushing:2011lr}. In contrast, substellar objects in star forming regions have much higher luminosities and therefore, in addition to many BDs, even isolated PMOs or planetary mass companion candidates have been identified in such regions \citep{1998Sci...282.1095T,1999ApJ...526..336O,2000MNRAS.314..858L,2000Sci...290..103Z,Neuhauser:2008fk,Kuzuhara:2011lr}. In order to reveal their formation processes, it is important to uncover their properties, such as binarity, space distribution, and especially initial mass function (IMF) in star forming regions. \\
\indent
Recent surveys of low-mass BDs and PMOs have been conducted in nearby star forming regions. In $\sigma$ Orionis, which is a well-studied field of PMOs \citep[e.g.,][]{Caballero:2007lr,2009A&A...506.1169B}, \citet{Pena-Ramirez:2012fk} have conducted a wide ($\sim 0.78\ {\rm deg}^2$) and deep (translated masses down to $\sim 0.004\ M_{\odot}$) optical/infrared imaging based on VISTA Orion survey data \citep{Petr-Gotzens:2011qy}. They have reported that $\sigma$ Orionis cluster harbors as many BDs (69 sources, $0.012-0.072\ M_{\odot}$) and PMOs (37 sources, $0.004-0.012\ M_{\odot}$) as low-mass stars (104 sources, $0.072-0.25\ M_{\odot}$). In the NGC1333, $\rho$ Ophiuchi and Chamaeleon-I cores, a systematic survey, called the Substellar Objects in Nearby Young Clusters (SONYC) project, was conducted \citep{2009ApJ...702..805S,2011ApJ...726...23G,2011ApJ...732...86M,2012ApJ...744..134M}. \citet{2012ApJ...744....6S} performed multi-object spectroscopy to extend their previous work over more targets by using Subaru/FMOS in NGC1333. They identified an L-type cluster member and estimated its mass to be as low as $0.006\ M_{\odot}$. However, since the construction of an IMF needs a large number of photometric and spectroscopic observations, current surveys in star forming regions are incomplete in the substellar regime, particularly in the planetary-mass domain. In order to understand the formation scenario of low-mass BDs and PMOs, we need to conduct deeper surveys for more targets in these star forming regions. \\
\indent
The Orion nebular cluster (ONC) is the most suitable star forming region for IMF studies. It is nearby, $\sim 450$ pc \citep{2008hsf1.book..483M}, and young, $\leq  $1 Myr \citep{1997AJ....113.1733H,2007MNRAS.381.1077R}. Previous studies of the ONC have concentrated on the central part of the ONC region. \citet{1997AJ....113.1733H} conducted an optical imaging survey and spectroscopic follow-up, including spectra for $\sim 900$ stars, and measured the stellar masses down to $\sim 0.1\ M_{\odot}$ within $4.5\ {\rm pc}\times 4.8\ {\rm pc}$  ($34' \times 36'$) of the Trapezium cluster. \citet{2000ApJ...540.1016L} conducted infrared imaging at the central region of the Trapezium cluster ($140'' \times 140''; 0.3\ {\rm pc} \times 0.3\ {\rm pc}$) and $K-$ band spectroscopy for $\sim 100$ sources. \citet{2004ApJ...610.1045S} performed 97 $J$ and $K-$band spectroscopy measurements of the inner $5.1' \times 5.1'$ region of the ONC \citep{2000ApJ...540..236H}. Although they confirmed many low-mass cluster members and even BDs of masses down to 0.02 $M_{\odot}$, the bright nebulosities make it difficult to extend the measurements to lower masses. In order to cover a larger area of the ONC, deep and wide imaging surveys have been conducted \citep{2010AJ....139..950R,2011AA...534A..10A,Da-Rio:2012lr}. However, no spectroscopic observations of the targets have been made yet. \citet{2005MNRAS.361..211L} have made observations focused on the outer regions of the ONC, where fainter nebulosities enable observations of fainter objects. They detected 33 faint planetary mass candidates. To confirm their cluster membership, spectroscopic observations were conducted and about 10 PMO candidates were identified as young stellar objects \citep{2006MNRAS.373L..60L,2009MNRAS.392..817W}. However, since many of the photometric candidates have not been spectroscopically examined yet, it is important to observe the candidates for the characterization of BDs and PMOs in this region. We therefore have conducted spectroscopic follow-up observations of the \citet{2005MNRAS.361..211L} candidates, and found two new young BDs and a BD/Planetary-mass boundary objects.\\
\indent
In section 2, we explain our observations and data reduction. In section 3, we show the analysis and results. In section 4, we present the HR-diagram and the number ratio of the stellar to substellar mass objects and discuss our conclusions.
\section{Observations and data reduction}
\subsection{Near-infrared spectroscopy}
Near infrared spectra of BD candidates in the ONC were obtained through two observing programs. The first data set was acquired on November 30, 2007, using the Multi-Object InfraRed Camera and Spectrograph (MOIRCS; Suzuki et al. 2008\nocite{2008PASJ...60.1347S}) mounted on the 8.2-m Subaru telescope (f/12.2 at the cassegrain focus). In this instrument, a $4' \times 7'$ field of view is covered by two HAWAII $2048 \times 2048$ HgCdTe arrays with a pixel scale of 0.117 arcsec/pixel.
 The second data set was obtained during December 3--7, 2010, using ISLE \citep{2006SPIE.6269E.118Y,Yanagisawa:2008lr}, a near-infrared imager and spectrograph for the cassegrain focus (f/18) of the 1.88-m telescope at Okayama Astronomical Observatory (OAO) of the National Astronomical Observatory of Japan (NAOJ). The detector is a HAWAII $1024 \times 1024$ HgCdTe array that covers a $4.3' \times 4.3'$ field of view with a pixel scale of 0.25 arcsec/pixel.\\
\indent
Target sources for the spectroscopy were selected from the catalogue of \citet{2005MNRAS.361..211L}, which presents the $JHK$ photometry of the BD candidates obtained by Gemini-South/Flamingos. The selection criteria were as follows. (1) All sources have prospective masses ranging from the hydrogen-burning limit ($<$ 0.075 $M_{\odot}$) down to the deuterium-burning limit of $0.013M_{\odot}$. (2) Sources that were close to or embedded in the bright nebulosity were avoided. For optimum sensitivity, objects located in areas of faint nebulosities were chosen (Figure 1 in \citet{2005MNRAS.361..211L}). (3) Sources were only chosen if they possessed fairly low visual absorption ($A_V < 7.5$). Although most of these objects have uncertainties ($\sim0.1$ mag) in their magnitude, two of them (208-736 and 215-652) have large errors ($> 0.2$ mag) in their $J$ and $H$ magnitudes. \citet{2010AJ....139..950R} have also observed these objects with ISPI, the facility infrared camera at the CTIO Blanco 4-m telescope, and have achieved better quality for 208-736 in $H-$band, and 215-652 in $J-$ and $H-$band. Therefore, in our study we used the magnitudes of \citet{2010AJ....139..950R} instead of \citet{2005MNRAS.361..211L} for these objects:  $H=17.37\pm0.10$ for 208-736 and $J=17.87\pm0.17$, $H=17.23\pm0.09$ for 215-652.\\
\indent
Unfortunately, we could only observe relatively bright candidates because auto-guider problems and high-humidity conditions prevented a full observing program with MOIRCS. Therefore, we focus on the good signal-to-noise ratio (S/N) candidates for MOIRCS in this paper. In total, 12 BD candidates were obtained, with eight of these objects obtained by MOIRCS and four obtained by ISLE (Table \ref{moslog}).\\
\indent
We took simultaneous spectra of eight of the BD candidates using the multi-object spectroscopy (MOS) mode of MOIRCS. We used a 1-arcsec slit width and the $HK500$ grism, optimized for 1.4--2.5 \mm \ and low-resolution ($R\sim 500$) spectra. To subtract the background noise, the objects were observed with slit nodding. Dome flats were obtained before a run. In order to calibrate telluric absorptions, we observed a standard star with an F-type spectrum (HD7386) in a similar air mass as the main observations for BDs.\\
\indent
The ISLE observations were carried out in the long-slit mode. The slit width was 1 arcsec and nod dithering was performed. A low resolution mode was selected with $R\sim 350$ in the $H-$band and $R\sim 450$ in the $K-$band. Observations on Dec 3 and Dec 4 were photometric. The observation of 065-207 on Dec 7 was taken under a condition of thin clouds and we carefully analyzed this source to reduce the effect of this condition. We employed a 1.5-arcsec slit width to acquire sufficient S/N. We obtained the dome flat before and after the observation. We selected an F-type star (HD24635) and a B-type star (HD34748) to measure the telluric absorption. When we use a B-type star possessing primarily strong hydrogen recombination lines as a telluric standard star, we should carefully deal with the intrinsic stellar lines since such lines cause artificial features in the science spectra. Therefore, we removed the intrinsic lines in the B-type stellar spectra by interpolating it with the neighboring flux.

\begin{table*}
	\begin{center}
		\caption{Summary of observed objects} \label{moslog}
		\begin{tabular}{lcccccccc}
			\hline
			\multicolumn{1}{c}{}			&		&$J$		&$H$	&$K$	&	&Obs time		&			&\\
			\multicolumn{1}{c}{Object}	&Data Set	&[mag]	&[mag]	&[mag]	&\av	&$H$,$K$[min]	&Reference\footnotemark[a]	&ID(LRT05)\footnotemark[b]\\
			\hline
			030-524&MOIRCS&17.88&17.26&16.77&1.8 \plms 1.5&60,60&1,2&178\\
			037-628&MOIRCS&18.52&17.66&16.99&4.4 \plms 1.5&60,60&2&94\\
			061-400&MOIRCS&18.31&17.33&16.48&5.7 \plms 1.5&60,60&2&270\\
			065-207&ISLE&13.47&12.81&12.24&0.8 $^{+1.5} _{-0.8}$&10,60&-&382\\
			072-638&ISLE&15.39&14.80&14.29&1.0 $^{+1.5} _{-1.0}$&10,-&1&70\\
			099-411&MOIRCS&17.08&16.40&15.99&2.3 \plms 1.5&60,60&-&256\\
			104-451&ISLE&15.10&13.99&13.13&5.8 \plms 1.5&5,30&-&230\\
			183-729&MOIRCS&18.05&17.38&17.24&2.3 \plms 1.5&60,60&1,2&18\\
			208-736&MOIRCS&18.28\footnotemark[c]&17.37\footnotemark[d]&16.56&5.0 \plms 2.4&60,60&-&8\\
			215-652&MOIRCS&17.87\footnotemark[d]&17.23\footnotemark[d]&17.15&2.0 \plms 2.0&60,60&-&56\\
			216-540&MOIRCS&16.95&15.89&15.47&3.7 \plms 1.5&60,60&-&155\\
			217-653&ISLE&14.99&14.26&13.32&2.4 \plms 1.5&5,30&1&55\\
			\hline
			\multicolumn{9}{@{}l@{}}{\hbox to 0pt{\parbox{150mm}{\footnotesize
				Note. Object names are coordinate based, following \citet{1996AJ....111..846O}. 030-524 means that the coordinate of the object is (R.A., Decl.) = (\timeform{05h35d03.0s}, \timeform{-05D25'24''}). $JHK$ magnitudes are taken from \citet{2005MNRAS.361..211L} with the exception of 208-736 and 215-652. Visual absorption ($A_V$) is estimated from a comparison between observed and synthetic $J$--$H$ vs $H$ color-magnitude diagrams.
				\par \noindent
				\footnotemark[a] The previous spectroscopic studies. 1. \citet{2007MNRAS.381.1077R}; 2. \citet{2009MNRAS.392..817W}
				\par \noindent
				\footnotemark[b] The running number is used in \citet{2005MNRAS.361..211L} 
				\par \noindent
				\footnotemark[c] The magnitude has a large uncertainty ($>0.2$ mag)
				\par \noindent
				\footnotemark[d] The magnitude is taken from \citet{2010AJ....139..950R} instead of \citet{2005MNRAS.361..211L}
			}\hss}}
		\end{tabular}
	\end{center}
\end{table*}

\subsection{Data reduction}
Both the multiple and long slit data were reduced using IRAF software. Standard techniques were used to reduce the MOIRCS multiple slit data. In order to subtract the sky background, exposures for each BD candidate pair were separated into their relative nod positions (A or B). We subtract the adjacent B frame from the A frame to decrease the residual. Flat fielding was corrected by using the dome flat frames. Note that channel 1 of MOIRCS was the engineering detector from Oct 2007 to Jun 2008 and had prominent, large ring-like high dark noise. Therefore we masked and did not use these regions in this data reduction. Cosmic ray cleaning was performed by L.A.Cosmic (Laplacian Cosmic Ray Identification; van Dokkum 2001\nocite{2001PASP..113.1420V}). Array distortion was corrected by the MCSGEOCORR task in the MCSRED\footnote{MCSRED was provided by Dr. Ichi Tanaka. See \url{http://www.naoj.org/staff/ichi/MCSRED/mcsred.html} software for MOIRCS imaging.}. To calibrate the wavelength of the data, OH lines were identified as a function of wavelength. After performing 5-pixel binning along the wavelength directions of the data to improve the S/N ratio, we extracted the spectra with the APALL task in IRAF. We correct spectral contamination due to telluric absorption using the standard star measurements. Extracted pairs from each nod position were combined. Finally, a de-reddening procedure was conducted with an adopted reddening parameter of 3.1. We used the $J$--$H$ vs $H$ color and magnitude of the objects and synthetic color-magnitude diagram \citep{2010arXiv1011.5405A} with age $\sim 1$ Myr and distance $\sim 450$ pc to derive the \av \ values using the reddening law of \citet{1989ApJ...345..245C}. \\
\indent
For the ISLE spectra, the reduction process was similar to that of MOIRCS. However, we did not correct the array distortion because the influence of the distortion was negligible. We obtained the $H-$band and $K-$band spectra separately.  Therefore, the $H-$band and $K-$band spectra were scaled to the 2MASS magnitude of the objects whose Vega magnitude and zero point magnitude were based on \citet{2003AJ....126.1090C}. The advantages of using 2MASS magnitudes are that the 2MASS magnitudes of our objects were obtained with low errors ($<$ 0.1 mag), and 2MASS $Ks-$band filter reduced the noise contribution from the thermal background beyond 2.3 \mm.
\section{Analysis and results}
To derive the effective temperature and surface gravity, we compare our observed spectra to a grid of synthetic spectra and find the best-fitting model in the $\chi ^2$ sense \citep[e.g.,][]{2007ApJ...657.1064M}. First, we obtain the best-fitting parameters (\S\ref{sec:first}). We then conduct a Monte Carlo simulation in order to constrain the error range of parameter estimates  (\S\ref{sec:final}) \citep[e.g.,][]{2008ApJ...678.1372C,Bowler:2009lr,Aller:2013fk}. We restrict the range of stellar parameters in the fit to effective temperatures $1800\ {\rm K} \leq \teff \leq 4900\ {\rm K}$ and surface gravity $3.0 \leq {\rm log(g)} \leq 5.5$. Also, we use a solar metallicity as the average one of low-mass members in the ONC.

\begin{figure*}
	\begin{center}
		\FigureFile(160mm, 100mm){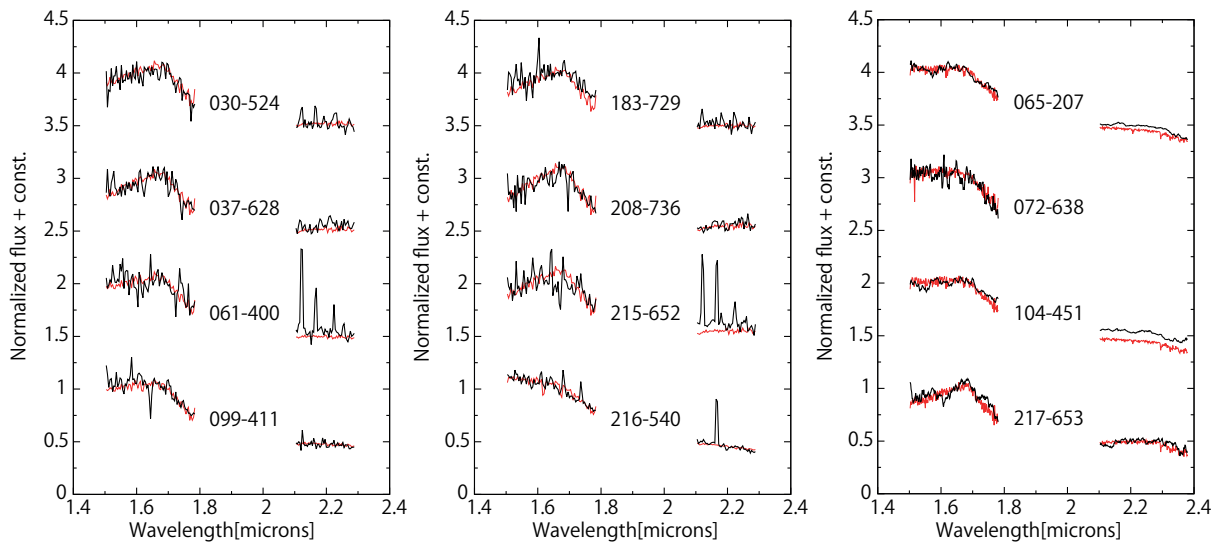}
	\end{center}
	\caption{Near infrared observed spectra. The left two panels show spectra obtained by MOIRCS and the right panel shows spectra obtained by ISLE. The observed (dereddened) spectra are shown as black solid lines while the best-fit model spectra are shown as red lines. The spectra of 061-400, 215-652 and 216-540 exhibit emission lines (H$_2$:2.12, 2.22$\mm$; Br$\gamma $: 2.16$\mm$).} \label{firstsp}
\end{figure*}

\subsection{BT-Settl model}
We use the BT-Settl model atmosphere \citep{2010arXiv1011.5405A}, which is available via the Phoenix web simulator. \citet{2010arXiv1011.5405A} updated the BT-Settl model  applying the revised solar abundances by \citet{2009ARA&A..47..481A}, the latest H$_2$O line list \citep{2008LPICo1405.8147B}, a cloud model based on condensation and sedimentation timescales by \citet{1978Icar...36....1R}, the supersaturation computed from pre-tabulated chemical equilibrium, and mixing from 2D radiation hydrodynamic simulations by \citet{2010A&A...513A..19F}. \\
\indent
BDs have an atmosphere including water vapor which causes strong absorption in the infrared regime. Because young BDs found in Trapezium are still gravitationally contracting, they have lower surface gravities than field dwarfs. Collision induced H$_2$ absorption (H$_2$-CIA) over the near infrared wavelength regime is pronounced at environments of strong surface gravity. Therefore, young BDs have a triangular $H-$band shape due to deep H$_2$O and weak H$_2$-CIA absorptions \citep{2006ApJ...639.1120K}. 

\subsection{$\chi ^2 -fitting$}
\subsubsection{Procedure 1 - Seek best fitting parameters}\label{sec:first}
In order to determine best fitting parameters, we conduct $\chi ^2$-fitting between the observed and BT-Settl model spectra. The results are shown in the column 3 or 4 of Table \ref{first}. The parameter ranges of the effective temperature and surface gravity are $1800\ {\rm K} \leq T_{\rm eff} \leq 4900\ {\rm K}$ and $3.0 \leq {\rm log(g)} \leq 5.5$. The range of the surface gravity is predicted from BT-Settl evolutional tracks for an age ranging from 1 Myr to 1 Gyr. Although the determinations of the metal content in the Orion region reach controversial conclusions, we use a solar metallicity as the average one of low-mass members in the ONC based on \citet{DOrazi:2009fk}. We seek the best \teff\ and log(g) to minimize the reduced $\chi ^2$ value: 
\begin{eqnarray}
	\chi ^2=\frac{1}{N-2} \sum^N \biggl(\frac{C f_{\rm mod}-f_{\rm obs}}{\sigma _{\rm obs}}\biggr)^2 \label{chi}
\end{eqnarray}
where  $N$ is the bin number of the wavelength, $f_{\rm mod}$ is the flux of the model spectra, $f_{\rm obs}$ is the flux of the observed spectra, and $\sigma _{\rm obs}$ the error of the observed spectra. The error is derived from the standard deviation of 4 combined spectra. $C$ is the scaling coefficient of $f_{\rm mod}$ with $f_{\rm obs}$ and is derived from $d\chi ^2/dC=0$ for Equation \ref{chi}:
\begin{eqnarray}
	C=\frac{(\sum f_{\rm obs}f_{\rm mod} /{\sigma _{\rm obs}})^2}{\sum ({f_{\rm mod}}/{\sigma _{\rm obs}})^2}
\end{eqnarray}
In order to minimize telluric contamination from water vapor, we restrict the regions for the fit 1.50--1.78 \mm\ and 2.10--2.29 \mm\ in the case of MOIRCS, and 1.5--1.78 \mm\ and 2.10--2.38 \mm\ for the ISLE data. 061-400, 215-652 and 216-540 have emission lines in the $K-$band spectra (H$_2$: 2.12, 2.22 $\mm$; Br$\gamma $: 2.16 $\mm$), which make the $\chi ^2$ value worse when we include the $K-$band data in the fit. Therefore, we exclude the emission lines in the fitting process. Also, we can see spiky substructures in some $H-$band spectra. For example, there is the spiky bump around 1.7 \mm\ in the spectrum of 208-736, which could be caused by partly very low-S/N ($\sim 1\ \sigma$). Such substructure affected by poor S/N make $\chi ^2$ value worse. When both $H$ and $K$ spectra exist, we perform $\chi ^2$-fitting in two cases: using only the $H-$band and using both the $H-$ and $K-$bands. \\
\indent
After the fit, we adopt the best-fitting parameters as those that minimize the $\chi ^2$ value. We show the selected wavelength ($H$ or $HK$) resulting in the best-fitting in column 5 of Table \ref{first}. In many cases we use both the $H-$ and $K-$bands to increase the sampling points, except for 065-207, 072-638 and 104-451. We adopt the result of fitting only using $H-$band for 072-638 due to a lack of $K-$band spectrum. Also, we employ the results of $H-$band for 065-207 and 104-451 because, in the fitting by only the $H-$band, we find a systematic offset between the observed and synthetic $K-$band spectra. The offset may be explained by the $K-$band excess from a circumstellar structure. Since the excess due to the circustellar structure is more apparent in the longer wavelength, we check $L'-$ band data of the objects. \citet{Lada:2004lr} presented the $L'-$band data, $L'=11.69$ for 065-207 and $L'=12.27$ for 104-451, by using VLT/ISAAC, which enable us to derive the dereddened photometric colors $H-L'=1.02$ and $H-L'=0.99$ respectively. Then, we estimate the synthetic colors $H-L'$ with the theoretical spectra fitting with $H-$band for 065-207 and 104-451, and with the $H$ and $L'$ filter response curves, which results in $H-L'=0.71$ (065-207) and $H-L'=0.70$ (104-451). The photometric colors of these objects are adequately redder than the synthetic colors ($\Delta (H-L') > 0.3$), which is strong evidence of a circumstellar structure. 

\subsubsection{Procedure 2 - Estimate the uncertainty}\label{sec:final}
We perform a Monte Carlo simulation, the parametric bootstrapping, to find the $T_{\rm eff}$ distribution, allowing us to determine the error range of $T_{\rm eff}$. In the simulation, we conduct a lot of $\chi ^2-$fittings between model spectra and randomly generated mock spectrum. There is an uncertainty in the parameter estimate from the $\chi ^2-$fitting due to both the S/N of the observed spectra and the uncertainty of visual absorption ($A_V$) estimate. Therefore, we separately evaluate the impact of these uncertainties on the derived $T_{\rm eff}$. We show the values of $T_{\rm eff}$ and errors in column 6 of Table \ref{first}. The procedures for deriving these values are as follows: \\
(1) In order to conduct a $\chi ^2$-fitting, we simulate a mock spectrum by adding an artificial noise to the theoretical spectrum. The parameters of the theoretical spectrum are determined based on the values derived in \S\ref{sec:first} and described in column 3 or 4 of Table \ref{first}. For example, in the case of 030-524, we use the theoretical spectrum with $T_{\rm eff}=2600$ K and log(g)=3.0. The artificial noise is randomly generated based on the error of observed spectrum, which means that we model the artificial noise at each wavelength as a drawn from a Gaussian distribution with a standard deviation adjusted to the 1$\sigma$ error of the observed spectrum at the corresponding wavelength. Note that this procedure to simulate the mock spectra differs from the previous studies \citep[e.g.,][]{2008ApJ...678.1372C} which simulated their mock spectra by adding artificial noise to the observed spectra.\\
(2) We determine the best-fitting parameters for the mock spectrum using $\chi ^2$ fitting.\\
(3) We repeat 10,000 iterations of {\it 1 -- 2} steps, making the distribution of $T_{\rm eff}$. We fit a Gaussian function to the distribution of $T_{\rm eff}$ measured in step {\it 2}. The error range of $T_{\rm eff}$ should be treated carefully since the error range is affected by the S/N of the spectrum and the uncertainty of the visual absorption $A_V$. By taking into account of the S/N of the spectrum, we adopt the 1 $\sigma$ confidence range of the Gaussian distribution as the error range $\Delta T_{\rm eff, S/N}$.\\
(4) In order to evaluate the impact of the uncertainty $\Delta A_V$ on the error range of $T_{\rm eff}$, we conduct the $\chi ^2-$fitting using three different \av;  the average value $A_V$, the upper limit $A_V + \Delta A_V$ and the lower limit $A_V - \Delta A_V$. For example, in the case of 030-524, we use three spectra dereddened with $(A_V - \Delta A_V, A_V, A_V+\Delta A_V)=(0.3, 1.8, 3.3)$. For the three spectra, we conduct the $\chi ^2-$fitting as described in \S\ref{sec:first} to seek the best fitting parameters.\\
(5) We obtain three effective temperatures: $T_{\rm eff}(A_V)$, $T_{\rm eff}(A_V+\Delta A_V)$ and $T_{\rm eff}(A_V-\Delta A_V)$. In order to describe the error affected by $\Delta A_V$, we define the upper error value as $\Delta T_{\rm eff, A_V, +}\equiv T_{\rm eff}(A_V+\Delta A_V)-T_{\rm eff}(A_V)$, and the lower error value as $\Delta T_{\rm eff, A_V, -}\equiv T_{\rm eff}(A_V)-T_{\rm eff}(A_V-\Delta A_V)$.\\
(6) We choose the larger of $\Delta T_{\rm eff, S/N}$ and $\Delta T_{\rm eff, A_V,\plms}$ as our $T_{\rm eff}$ uncertainty. In most cases, our error is dominated by the uncertainty in $A_V$.\\

\subsubsection{Results}\label{sec:result}
The obtained spectra and best-fitting theoretical spectra from \S\ref{sec:final} are shown in Figure.\ref{firstsp}, and the estimated physical parameters are shown in Table \ref{first}. Most objects are well reproduced by the BT-Settl models. However, 216-540 required a change to the extinction value $\Delta \av \sim 2$ to obtain better $\chi ^2$ values, which results in $\teff =3320^{+760} _{-270}$ K.\\
\indent
Although some objects have slightly higher $\teff$ than those estimated by the previous works (037-628 and 188-739), our result is largely consistent with those of previous works \citep{2007MNRAS.381.1077R,2009MNRAS.392..817W}. Note however that \teff $=2750^{+250} _{-170}$ K of 061-400 is much higher than the previous result $\teff \leq 2400$ K \citep{2009MNRAS.392..817W}. In \citet{2009MNRAS.392..817W}, the photometry and spectroscopy for this object were originally based on \citet{2001MNRAS.326..695L}, in which they adopted $J=18.250$ and $H=17.706$. If we adopt the same magnitudes, we obtain $A_V=0$, which is lower than the adopted value $A_V=5.7$ in this work, and the lower $A_V$ results in a lower temperature. However, \citet{2010AJ....139..950R} provided a magnitude of $J=18.35\pm0.20$ mag and $H=17.46\pm0.10$ mag. Our adopted magnitudes are consistent with these more recent values. Thus, we use our result for 061-400 in the following discussions.

\section{Discussion}
Our observations identify new very-low-mass objects in the Orion nebular cluster. Here we plot $\teff$ and absolute $H$ magnitude on a synthetic HR diagram. As a result, we obtain three new young BD candidates: 099-411 is an object with boundary mass between stellar and substellar mass, 215-652 is a BD candidate with emission lines associated with youth, and 208-736 is a very low mass object which is even a BD/Planetary-mass object. In this section we will discuss the properties of each object based on the HR diagram, and then derive a number ratio of the stellar to substellar mass objects and discuss the star formation process in the outer region of the ONC.
\subsection{HR diagram}\label{sec:hr}
\begin{figure*}
	\begin{center}
		\FigureFile(80mm, 50mm){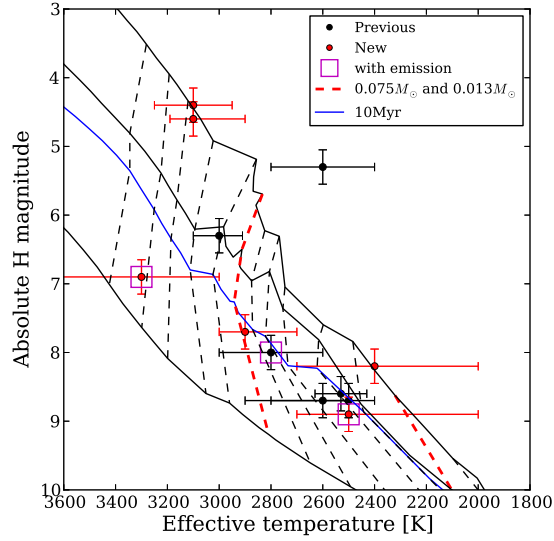}
	\end{center}
	\caption{\teff\ vs absolute $H-$band HR diagram for 12 objects in our spectroscopic sample. The lines indicate the synthetic HR-diagram \citep[BT-Settl;][]{2010arXiv1011.5405A}. The solid lines show the age isochrones 1, 5, 10, 100 Myr from top to bottom. The dashed lines show the mass isochrones 0.007, 0.013, 0.02, 0.03, 0.04, 0.05, 0.06, 0.075, 0.1, 0.11, 0.15, 0.2, 0.3 $M_{\odot}$ from right to left (the red dashed lines indicate 0.013 and 0.075 $M_{\odot}$). The red dots mean newly observed objects and the black dots mean follow-up objects from previous works \citep{2007MNRAS.381.1077R,2009MNRAS.392..817W}. The magenta squares represent the objects with emission lines.}\label{HR}
\end{figure*}

We estimate the masses of the BD candidates by plotting the effective temperature and the absolute $H$ magnitude on an HR diagram overlaid with theoretical isochrones (Figure \ref{HR}). The error range of the mass corresponds to the error of the effective temperature in the HR diagram. We show the derived masses and the errors in Table \ref{first}. In order to convert the apparent magnitudes to the absolute magnitudes, we employ 450 pc (DM = 8.26) as the ONC distance. This value is the average distance in the previous investigation \citep{2008hsf1.book..483M}. The error of the distance is at most 50 pc (DM = 0.25) but the error is not significant compared with the error of the effective temperature in deriving the mass. \\
\indent
In Figure \ref{HR}, it appears that there are several old age objects ($\sim 10$ Myr) although the ONC is a very young cluster, $\leq 1$ Myr \citep{1997AJ....113.1733H}. In order to explain the old population, we consider several scenarios. The first is that we are seeing contamination from another foreground population. It is known that the ONC is neighboured by several group of stars. \citet{Alves:2012lr} have investigated the surrounding population of the ONC, and reported the contiguous population, NGC 1980, overlaps significantly with the ONC. Since they have shown that NGC 1980 can make up for more than $10-20$ \% of the ONC population, the projection of the foreground population may generate the age dispersion. However, its members are thought to be only $\sim 4-5$ Myr and therefore cannot account for the 10 Myr population on the HR diagram. Second we consider the case that the older objects are foreground galactic field stars. In particular, it is standing that 030-524 and 216-540 are older than 10 Myr. Since the local number density peaks at M3-M4 \citep{2008ASPC..393...51C}, such objects are often contaminated. In contrast, it seems difficult to be contaminated by fainter objects since the local number density steeply decreases for the later spectral type ($\geq $\rm M5), and the dense background molecular clouds prohibit penetration of the light of the very cool stars. In order to confirm how frequently field stars can contaminate our sample,  we calculate the probability of the contamination in the spectral type range of our interest.\\
\indent
Among 12 objects observed in our work, eight of those have $2400\:{\rm K} < T <  3000\:{\rm K}$, corresponding from M9 to M6 \citep{2003ApJ...593.1093L}. First, we estimate the number of contaminating field stars with the spectral type M6 to M9. For the purpose, we adopt the same method as \citet{2008A&A...488..181C} with the FOV of MOIRCS $\sim 4' \times 7'$\footnote{Since \citet{2008A&A...488..181C} did not provide information for the $H-$band, we quote the $J$--$H$ color from \citet{2008AJ....135..785W}.}. We conservatively adopt a magnitude range of $12 < H < 18$ to include the brightest magnitude $H=12.8$  mag (065-207) and the faintest magnitude $H=17.7$ mag (037-628) in our work. For the ONC, we use $E_{B-V}\sim 0.32$ from \citet{1968ApJ...152..913L}. As a result, the number of contaminating field stars is 0.23.\\
 \indent
We estimate the probability of contamination by using the Poisson distribution with a mean of 0.23. The probability of detecting at least one contaminant is 20.3\%. This means that one field BD having the similar temperature to 030-524 may be observed in our small field, and thus the number of contaminants is not significant for M6--M9 dwarfs. However, when M3--M5 dwarfs are included, the expected number of contaminants increases to 11.5, so the probability of early M contaminant with a similar temperature to 216-540 is quite high. Therefore, we cannot make a strong statement about the membership of the M3--M5 objects from the contaminant probability.\\
\indent
Next, we verify if the objects which seem to be older in photometry have any evidence of youth. Since protostellar objects are still experiencing gravitational contraction, they have a low surface gravity. 030-524 has a sharp triangular $H-$band spectrum, which is evidence of low gravity and therefore a protostar undergoing gravitational contraction. 216-540 shows a Br$\gamma $ emission in the $K-$band, which is associated with the mass accretion process and known as a common mass accretion tracer \citep{Muzerolle:1998lr}. 061-400 and 215-652 also have an emission line in the $K-$band spectra, which means that  these objects are mass accreting protostars. As has been noted, several objects have spectral features consistent with being young objects although they seem to be old on our HR diagram. One explanation for this discrepancy is that these objects are protostars with edge-on disks which scatter light from the central star and decrease its apparent magnitude. In fact,  the HST/ACS study of \citet{Ricci:2008lr} reported that 037-628, 061-400, 215-652 and 216-540 have ionized disks seen in emission. The existence of such objects has been indicated by previous surveys not only in the ONC \citep{2004ApJ...610.1045S} but also in other regions \citep[e.g.,][]{2002ApJ...580..317B,2003ApJ...593.1093L}. For example,  in two T Tauri systems HH30 and HK Tau C with edge-on circumstellar disks, the $K-$band magnitude is more than 3 mag lower than expected for their age, distance and spectral type \citep{Burrows:1996qy,Stapelfeldt:1998lr}. Consequently, the apparently older objects can be explained as young stellar objects with edge-on disks. Also, the episodic accretion process may explain the low luminosity of the apparently older objects, even without edge-on disks \citep{Baraffe:2009fk}. However, in order to reinforce the validity of their membership, we need a further evidence of youth. For example, we should investigate the infrared excess due to the circumstellar disks at longer wavelengths.

\begin{table*}
	\begin{center}
		\caption{The result of $\chi ^2-$fitting}\label{first}
		\begin{tabular}{lcccclcccc}
			\hline
			\multicolumn{1}{c}{}           &                                          &$H$\footnotemark[b]                 &$HK$\footnotemark[b]             &                                &			&&Reference&Reference\\
			\multicolumn{1}{c}{Name}&$M_H$\footnotemark[a]&$T_{\rm eff} $ log(g) $\chi ^2$&$T_{\rm eff}$ log(g) $\chi ^2$&Fit\footnotemark[c]&$T_{\rm eff}$ [K]\footnotemark[d]&Mass [$M_{\odot}$]\footnotemark[e]&SpT\footnotemark[f]&$T_{\rm eff}$ [K]\footnotemark[g]\\
			\hline
			030-524&8.7&3000 3.0 0.9&2600 3.0 2.1&$HK$&$2600^{+300} _{-100}$&$0.037^{+0.049} _{-0.015}$&M8.0\plms 0.75&$2710^{+127} _{-233}$\\
			037-628&8.6&2700 3.5 1.7&2500 3.0 1.7&$HK$&$2500^{+100} _{-100}$&$0.022^{+0.006} _{-0.005}$&$>$M9.5&$<2400$\\
			061-400&8.0&3400 3.0 1.8&2800 3.5 3.0&$HK$&$2800^{+200} _{-200}$&$0.046^{+0.056} _{-0.018}$&$>$M9.5&$<2400$\\
			065-207&4.4&3100 4.5 0.3&2900 4.0 1.0&$H$&$3100^{+150} _{-150}$&$0.145^{+0.97} _{-0.039}$&-&-\\
			072-638&6.3&3000 5.5 0.9&- - -&$H$&$3000^{+100} _{-90}$&$0.097^{+0.018} _{-0.019}$&M6.5\plms 0.5&2935\plms 55\\
			099-411&7.7&2900 4.5 4.8&2900 4.5 3.1&$HK$&$2900^{+100} _{-200}$&$0.062^{+0.030} _{-0.026}$&-&-\\
			104-451&4.6&3100 4.0 1.0&2500 3.0 2.8&$H$&$3100^{+90} _{-200}$&$0.134^{+0.078} _{-0.031}$&-&-\\
			183-729&8.7&2800 3.0 1.9&2500 3.0 2.0&$HK$&$2500^{+300} _{-100}$&$0.022^{+0.035} _{-0.006}$&$\geq $M9.5&$<2400$\\
			208-736&8.2&2500 3.0 1.4&2400 3.0 1.7&$HK$&$2400^{+300} _{-400}$&$0.016^{+0.022} _{-0.010}$&-&-\\
			215-652&8.9&2900 3.5 2.8&2500 3.0 4.2&$HK$&$2500^{+200} _{-500}$&$0.022^{+0.027} _{-0.016}$&-&-\\
			216-540&6.9&3300 5.0 3.8&3300 5.0 2.6&$HK$&$3300^{+700} _{-300}$&$>0.100$&-&-\\
			217-653&5.3&2900 3.0 2.3&2600 3.0 1.7&$HK$&$2600^{+200} _{-200}$&$0.031^{+0.028} _{-0.015}$&M7.75\plms 0.5&$2752^{+80} _{-120}$\\
			\hline
			\multicolumn{4}{@{}l@{}}{\hbox to 0pt{\parbox{160mm}{\footnotesize
				\footnotemark[a] Absolute $M_H$ magnitudes.
				\par\noindent
				\footnotemark[b] The values are derived from $\chi ^2$-fitting for seeking best fitting parameters (see \S\ref{sec:first}). The $T_{\rm eff}$ and log(g) are used for simulating the mock spectrum (see \S\ref{sec:final}).
				\par\noindent
				\footnotemark[c] This column shows the wavelength used for determine best fitting parameters (see \S\ref{sec:first}).
				\par\noindent
				\footnotemark[d] The best $T_{\rm eff}$ and the uncertainty which is derived from the Monte Carlo simulation (see \S\ref{sec:final}).
				\par\noindent				
				\footnotemark[e] Masses derived from the HR diagram (see \S\ref{sec:hr} and Figure \ref{HR}).
				\par\noindent
				\footnotemark[f] The reference of spectral type (SpT) is shown in Table \ref{moslog}. When the object has been observed both in \citet{2007MNRAS.381.1077R} and \citet{2009MNRAS.392..817W}, we use the SpT of  \citet{2009MNRAS.392..817W}.
				\par\noindent
				\footnotemark[g] These quantities were derived using the reference spectral type to effective temperature scale of \citet{2003ApJ...593.1093L}
			}\hss}}
		\end{tabular}
	\end{center}
\end{table*}

\subsection{Stellar to substellar mass number ratio}\label{sec:NumberRatio}
Here, we derive the stellar to substellar mass number ratio $R$, which can be a clue to discuss a possibility of difference in the formation process between stars and substellar objects \citep[e.g.,][]{2003ApJ...593.1093L,2008ApJ...683L.183A}. In previous works, there are two definitions of $R$, which are given as,
\begin{eqnarray}
R_1&=&\frac{N(0.08<M/M_{\odot}<1)}{N(0.03<M/M_{\odot}<0.08)}\\
R_2&=&\frac{N(0.08<M/M_{\odot}<10)}{N(0.02<M/M_{\odot}<0.08)}
\end{eqnarray}
In this paper, we derive $R$ following the above definitions. We measure the $R_1$ and $R_2$ according to the following steps. (1) We choose objects with $A_V < 7.5$ out of \citet{2005MNRAS.361..211L} whose observation was conducted for the $1.7'$--$5.7'$ regions from the ONC center. The completeness limit for $A_V$ is determined to be $A_V=7.5$. (2) We consider that objects with spectroscopically known properties are fiducial sample to investigate $R$. However, the sample which we spectroscopically observed is still insufficient for a statistical discussion. In order to increase the sample size, we have compiled information of spectroscopic observations of the corresponding $A_V < 7.5$ sample from previous studies \citep{1997AJ....113.1733H,2006MNRAS.373L..60L,2007MNRAS.381.1077R,2009MNRAS.392..817W}. Also, non-member stars are excluded from the sample based on the previous studies. \citet{1997AJ....113.1733H} conducted an optical imaging and spectroscopic survey for stellar mass objects in a large region of ONC, and almost 80\% complete down to $0.1\ M_{\odot}$. \citet[][infrared]{2006MNRAS.373L..60L}, \citet[][optical]{2007MNRAS.381.1077R} and \citet[][infrared]{2009MNRAS.392..817W} are spectroscopic studies for the substellar mass candidates ($< 0.075\ M_{\odot}$) found by \citet{2001MNRAS.326..695L} and \citet{2005MNRAS.361..211L}. (3) In order to avoid inconsistencies from different mass determinations, we recalculate the masses of individual samples using the HR-diagram as explained in \S\ref{sec:hr}, instead of the masses derived in the previous papers. (4) We correct for the bias due to sample incompleteness of the spectroscopic survey. As described above, \citet{1997AJ....113.1733H} observed stellar mass objects, but the other studies have concentrated on substellar mass objects. As a result, the spectroscopic completeness is 68\% in the stellar mass regime and 58\% in the substellar mass regimes as defined for the calculation of $R_1$. Also, the completeness is 72\% in the stellar mass regime and 55\% in the substellar mass regime used for $R_2$ . When we estimate the number ratio of the stellar to substellar mass, the bias of the completeness causes the number ratio to be overestimated. Therefore, in order to correct the bias, we add three hypothetical substellar objects to calculate $R_1$ and five hypothetical substellar objects to calculate $R_2$ according to the completeness of the stellar mass regime. Consequently, we determine $R_1=90/26=3.5\pm 0.8$ and $R_2=110/34=3.2^{+0.7} _{-0.6}$ with $1.7'$--$5.7'$ regions from the ONC center, and the errors are estimated using the method of \citet{2012ApJ...744....6S} who use the confidence interval provided by \citet{Cameron:2011lr}.
\begin{table*}
	\begin{center}
		\caption{Star to substellar number ratio in the orion nebular cluster} \label{NumberRatioTable}
		\begin{tabular}{lccc}
			\hline
			\multicolumn{1}{c}{reference}&$R_1$&$R_2$&$r$\\
			\hline
			This work							&3.5\footnotemark[a] (2.7-4.3)	&3.2\footnotemark[a] (2.6-3.9)	&$1.7'<r<5.7'$\\
			\citet{2003ApJ...593.1093L}			&-						&3.8						&$r<1.5'$\\
			\citet{2004ApJ...610.1045S}			&3.3\footnotemark[b] (2.6-4.1)	&5.0						&$r<2.5'$\\
			\citet{2011AA...534A..10A} inner		&7.2\ (1.6-12.8)	&-			&$2.9'<r<5.8'$\\
			\citet{2011AA...534A..10A} cluster		&2.4 (2.2-2.6)	&-			&$2.9'<r<12.5'$\\
			\hline
			\multicolumn{4}{@{}l@{}}{\hbox to 0pt{\parbox{160mm}{\footnotesize
			\footnotemark[a] The error range is estimated in the same manner as \citet{2012ApJ...744....6S}.
			\par\noindent
			\footnotemark[b] \citet{2008ApJ...683L.183A} have compiled the study of \citet{2004ApJ...610.1045S} and shown this value.
			}\hss}}
		\end{tabular}
	\end{center}
\end{table*}
\\
\indent
\citet{2012ApJ...744....6S} listed $R$ values for various star forming regions. In comparison with the other star forming regions, our derived $R_1$ and $R_2$ are second lowest values, which means that the number ratio of substellar mass objects in the ONC is relatively high compared with the other star forming regions. In the ONC, the $R$ has been investigated from its center to outer edge. \citet{2011AA...534A..10A} have discontinuously covered the large area ($26' \times 33'$) using HST/NICMOS Camera 3. In the total of the observed area, they have shown $R_1=2.4\pm0.2$ whose small value suggests a flat IMF in the substellar mass regime. $R_1({\rm Andersen}, cluster)$ is lower than our derived $R_1({\rm Suenaga})=3.5\pm 0.8$, which means that in a wide field of view the number of substellar objects is significantly high in the ONC. Meanwhile, \citet{Da-Rio:2012lr} have also covered a large field of view ($32' \times 33'$) with WFI camera mounted on the 2.2-m MPG/ESO telescope. They have constructed IMF down to 0.02 $M_{\odot}$ which declines steeply with decreasing mass, and they have found no evidence of IMF flattening toward the substellar mass regime. The inconsistency between \citet{2011AA...534A..10A} and \citet{Da-Rio:2012lr} may be explained by a different estimate of background contamination. In order to confirm the validity of these studies, we should extend the coverage area of our spectroscopic study with the multi-object spectrographs.\\
\indent
Considering the inner regions ($r<2.5'$) in the ONC, \citet{2003ApJ...593.1093L} have measured $R_2=3.8$. The properties of brighter objects ($K<12$) in their sample are taken from $K-$band spectroscopy, however the fainter substellar objects are mostly based on photometric data with HST/NICMOS camera 3 \citep{2000ApJ...540.1016L}. On the other hand, the inner regions was spectroscopically investigated by \citet{2004ApJ...610.1045S}, who has indicated $R_2({\rm Slesnick})=5.0$. \citet{2008ApJ...683L.183A} have compiled the study of \citet{2004ApJ...610.1045S}, and shown $R_1({\rm Slesnick})=3.3^{+0.8} _{-0.7}$ in this region. Our derived $R_1({\rm Suenaga})=3.5\pm0.8$ is consistent with the value $R_1({\rm Slesnick})$, but $R_2({\rm Suenaga})=3.2^{+0.7} _{-0.6}$ is lower than $R_2({\rm Slesnick})$. Since our observations focus on the outer regions of the ONC, in contrast to \citet{2004ApJ...610.1045S}, the central concentration of the high mass population ($>1 M_{\odot}$) found by \citet{Hillenbrand:1998lr} may explain the difference in measure $R_2$ values. \\
\indent
The outer regions ($2.9'<r<12.5'$) were observed by \citet{2011AA...534A..10A}. They reported that $R_1({\rm Andersen}, inner)=7.2\pm 5.6$ at the $2.9'$--$5.8'$ area is, although having the large error, consistent with the value $R_1({\rm Slesnick})$ of the inner part ($r<2.5'$). In addition, \citet{2011AA...534A..10A} found that the number ratio $R$ decreases as a function of radius. Our result suggests that $R_1({\rm Suenaga})=3.5\pm0.8$ of the outer part ($1.7'<r<5.7'$) is consistent with the $R_1({\rm Slesnick})$ of the inner area of the ONC ($r<2.5'$). This result suggests that the cluster distribution does not decrease simply as a function of radius. \\
\\
To summarize, we obtained 12 spectra with the multi-object slit and long-slit spectrograph. We confirm eight of those show strong water absorptions and have very low temperatures ($< 3000$ K). Two of cold objects are new young BDs and one of the objects is a BD/Planetary-mass boundary object. We estimate the stellar to substellar mass number ratio to be $R_1=3.5\pm 0.8$ and $R_2=3.2\pm 0.6$, the former is consistent with the value of the inner part $R_1$, but the latter is lower than the inner part $R_2$. Although we add 12 results of BD candidates to the spectroscopic sample, the follow-up observations of additional targets are ongoing. When the survey is completed, by the multi-object slit spectrograph, we can reveal the low mass end of the IMF.

\bigskip
We thank the assistance of the Subaru telescope and the Okayama Astronomical Observatory for allowing the use of their facilities. We appreciate an anonymous referee whose valuable comments and suggestions have improved the quality of the manuscript. We are grateful to L. C. Vargas for a careful reading of this paper and providing a lot of insightful comments. We are also thankful to F. Allard for allowing us access to her most recent calculation. This work is partly supported by the JSPS fund (22000005). T. S. acknowledges the support of a NINS Program for Cross-disciplinary Study. M.K. is financially supported by the JSPS through the JSPS Research Fellowship for Young Scientists (No. 25-8826). This publication makes use of data provided by SMOKA. A photometric catalogue is obtained from VizieR.

\bibliographystyle{aa}
\bibliography{suenaga2011}
\end{document}